\documentclass[journal=jacsat,manuscript=article]{achemso}

\usepackage[version=3]{mhchem} % Formula subscripts using \ce{}
\usepackage[utf8]{inputenc}
\usepackage[T1]{fontenc}
\usepackage{textcomp}
\usepackage{ulem}
\usepackage[separate-uncertainty = true, multi-part-units=single, range-phrase=--, range-units = single, detect-all]{siunitx}
\usepackage{tikz}
\author{Anas Elbaz}
\affiliation[C2N]
{Université Paris-Saclay, CNRS, C2N, 91120, Palaiseau, France}
\alsoaffiliation[ST]
{STMicroelectronics, Rue Jean Monnet 38054 Crolles, France}
\author{Riazul Arefin}
\affiliation[C2N]
{Université Paris-Saclay, CNRS, C2N, 91120, Palaiseau, France}
\author{Emilie Sakat}
\affiliation[C2N]
{Université Paris-Saclay, CNRS, C2N, 91120, Palaiseau, France}
\author{Binbin Wang}
\affiliation[C2N]
{Université Paris-Saclay, CNRS, C2N, 91120, Palaiseau, France}
\author{Etienne Herth}
\affiliation[C2N]
{Université Paris-Saclay, CNRS, C2N, 91120, Palaiseau, France}
\author{Gilles Patriarche}
\affiliation[C2N]
{Université Paris-Saclay, CNRS, C2N, 91120, Palaiseau, France}
\author{Antonino Foti}
\affiliation[PICM]
{LPICM, CNRS, Ecole Polytechnique, Université Paris-Saclay, 91120 Palaiseau, France}
\author{Razvigor Ossikovski}
\affiliation[PICM]
{LPICM, CNRS, Ecole Polytechnique, Université Paris-Saclay, 91120 Palaiseau, France}
\author{Sebastien Sauvage}
\affiliation[C2N]
{Université Paris-Saclay, CNRS, C2N, 91120, Palaiseau, France}
\author{Xavier Checoury}
\affiliation[C2N]
{Université Paris-Saclay, CNRS, C2N, 91120, Palaiseau, France}
\author{Konstantinos Pantzas}
\affiliation[C2N]
{Université Paris-Saclay, CNRS, C2N, 91120, Palaiseau, France}
\author{Isabelle Sagnes}
\affiliation[C2N]
{Université Paris-Saclay, CNRS, C2N, 91120, Palaiseau, France}
\author{Jérémie Chrétien}
\affiliation[CEA-INAC]
{Univ. Grenoble Alpes, CEA, IRIG-DePhy, 38054 Grenoble, France}
\author{Lara Casiez}
\affiliation[CEA-INAC]
{Univ. Grenoble Alpes, CEA, LETI, 38054 Grenoble, France}
\author{Mathieu Bertrand}
\affiliation[CEA-INAC]
{Univ. Grenoble Alpes, CEA, LETI, 38054 Grenoble, France}
\author{Vincent Calvo}
\affiliation[CEA-INAC]
{Univ. Grenoble Alpes, CEA, IRIG-DePhy, 38054 Grenoble, France}
\author{Nicolas Pauc}
\affiliation[CEA-INAC]
{Univ. Grenoble Alpes, CEA, IRIG-DePhy, 38054 Grenoble, France}
\author{Alexei Chelnokov}
\affiliation[leti]
{Univ. Grenoble Alpes, CEA, LETI, 38054 Grenoble, France}
\author{Philippe Boucaud}
\affiliation[CRHEA]
{Université Côte d’Azur, CNRS, CRHEA, Rue Bernard Grégory, CS 10269, 06905
Sophia-Antipolis Cedex France}
\author{Federic Boeuf}
\affiliation[STmicroelectronics]
{STMicroelectronics, Rue Jean Monnet 38054 Crolles, France}
\author{Vincent Reboud}
\affiliation[leti]
{Univ. Grenoble Alpes, CEA, LETI, 38054 Grenoble, France}
\author{Jean-Michel Hartmann}
\affiliation[leti]
{Univ. Grenoble Alpes, CEA, LETI, 38054 Grenoble, France}
\author{Moustafa El Kurdi}
\affiliation[C2N]
{Université Paris-Saclay, CNRS, C2N, 91120, Palaiseau, France}
\email{moustafa.el-kurdi@u-psud.fr}

\title[An \textsf{achemso} demo]
 {Reduced lasing thresholds in GeSn microdisk cavities with defect management of the optically active region}

\abbreviations{IR,NMR,UV}


\begin{document}

\begin{tocentry}

\includegraphics[width=\linewidth]{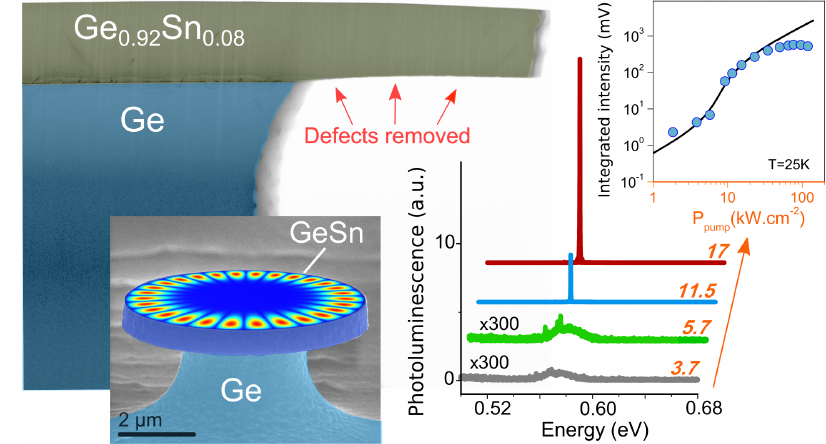}
\includegraphics[width=\linewidth]{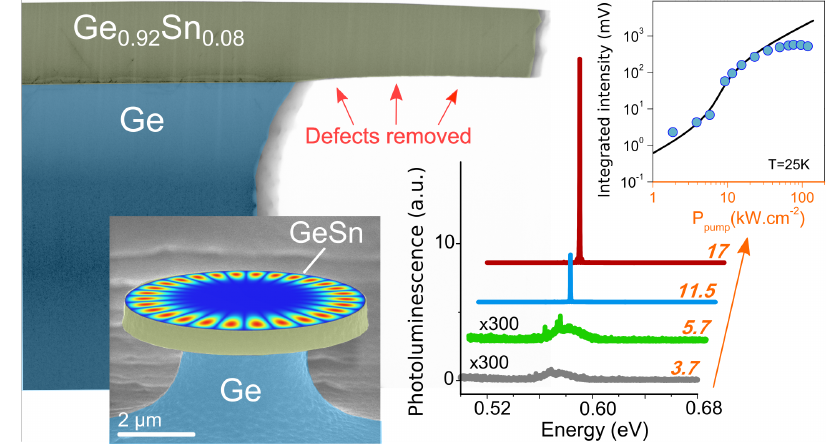}
\includegraphics[width=\linewidth]{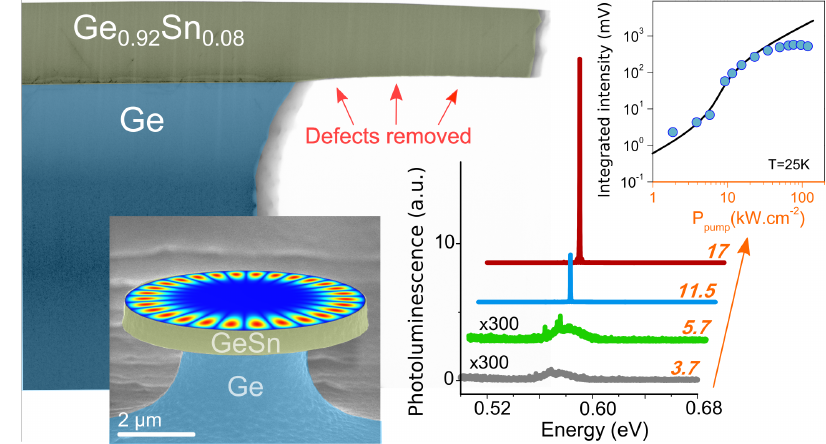}
Some journals require a graphical entry for the Table of Contents.
This should be laid out ``print ready'' so that the sizing of the
text is correct.

Inside the \texttt{tocentry} environment, the font used is Helvetica
8\,pt, as required by \emph{Journal of the American Chemical
Society}.

The surrounding frame is 9\,cm by 3.5\,cm, which is the maximum
permitted for \emph{Journal of the American Chemical Society}
graphical table of content entries. The box will not resize if the
content is too big: instead it will overflow the edge of the box.

This box and the associated title will always be printed on a
separate page at the end of the document.

\end{tocentry}


\begin{abstract}
GeSn alloys are nowadays considered as the most promising materials to build Group IV laser sources on silicon (Si) in a full complementary metal oxide semiconductor-compatible approach. Recent GeSn laser developments rely on increasing the band structure directness, by increasing the Sn content in thick GeSn layers grown on germanium (Ge) virtual substrates (VS) on Si. These lasers nonetheless suffer from a lack of defect management and from high threshold densities. In this work we examine the lasing characteristics of GeSn alloys with Sn contents ranging from 7 \% to 10.5 \%. The GeSn layers were patterned into suspended microdisk cavities with different diameters in the 4-\SI{8 }{\micro\meter} range. We evidence direct band gap in GeSn with 7 \% of Sn and lasing at 2-\SI{2.3 }{\micro\meter} wavelength under optical injection with reproducible lasing thresholds around \SI{10 }{\kilo\watt\per\square\centi\meter}, lower by one order of magnitude as compared to the literature. These results were obtained after the removal of the dense array of misfit dislocations in the active region of the GeSn microdisk cavities. The results offer new perspectives for future designs of GeSn-based laser sources. 
\end{abstract}

\section{Introduction}
The weak emission efficiency of group IV materials like Ge and Si can be traced back to their indirect band gap structure. 
One very promising option to overcome this limit relies on the alloying of Ge with Sn.\cite{wirths_lasing_2015,Soref2016} Inherently direct band gap materials can indeed be obtained from GeSn alloys with Sn contents around 7-8 \% and above. The material directness can be quantified by the energy splitting, $\Delta E_{L-\Gamma}=E_L-E_{\Gamma}$, between the indirect L valley and the direct valley ${\Gamma}$. This energy splitting increases with the Sn content in strain-free GeSn alloys.\cite{chen_increased_2011,grzybowski_next_2012} This very attractive feature has been the subject of numerous researches, the aim being to fabricate lasers compatible with group IV elements and complementary metal oxide semiconductor (CMOS) process\cite{du_study_2019,reboud_optically_2017,Thai:18} which has recently led to an electrically injected source.\cite{EpumpLaserGeSnArk} As the Sn solubility in Ge is limited to only 1 \%, i.e. well below the required one for direct band gap, one has to develop complex metastable growth process for homogeneous incorporation of Sn in Ge. 
The growth of high quality alloys still remains a challenging task. Growth temperatures below 400\textdegree C are mandatory, limiting the crystalline quality and the ability to cure defects by thermal annealing.\cite{zaumseil_thermal_2018} Moreover, the lattice mismatch between Ge-VS on Si and GeSn generates compressive strain in the as-grown layers. The compressive strain is known to counteract the effect of Sn incorporation, as it decreases the $E_L - E_\Gamma$ splitting energy and can even invert the band gap alignment.\cite{gupta_achieving_2013} One strategy to overcome this compressive stress while increasing the Sn content in the layer is to reach plastic relaxation. Several approaches were indeed developed to incorporate as much Sn as possible in GeSn alloys to increase their directness while managing the lattice mismatch between the Ge-VS and the epitaxially-grown GeSn layer. With the step graded layer technique \cite{Aubin2017,Dou2018,Dou2018a}, misfit dislocations are confined in the lower Sn content layers while high optical quality, high Sn content active layers lie on top. Increasing the Sn content results however in higher density of interface defects, which scales with the lattice mismatch and thus the Sn content.\cite{Gencarelli15032013} Consequently, a lower crystalline quality is inherent to high Sn content GeSn/Ge growth, which is in turn detrimental to carrier optical recombination dynamics \cite{pezzoliACSphotonics}. This would explain the very high pumping levels required to reach population inversion in previously reported high Sn content (>10 \%) GeSn lasers although active layers had a direct band gap.
One possibility to overcome this limitation is to confine carriers away from interfaces by using SiGeSn barriers in GeSn/SiGeSn double heterostructures or multi-quantum wells (MQW) stackings.\cite{vondenDriesch2018,stange_short-wave_2017,MQWSiGeSnSmall,SingSiGeSnQWFY}. A reduction of lasing threshold down to 40 kW cm$^{-2}$,\cite{StangeMQWlaser} as compared with hundreds of kW cm$^{-2}$ in bulk layers, was obtained with this approach. However MQWs suffer from rather small valence band offsets and reduced hole confinement. Their distinct advantages are thus conserved only up to 100 K\cite{StangeMQWlaser}.
\newline
 No works have so far focused to our knowledge on the deleterious impact that defects might have on lasing. To provide some insight about it, we studied relaxed GeSn microdisk cavities with attempts to suppress defects from their active region, specifically the dense array of misfit dislocations at the GeSn/Ge interface. One obvious way is to start with low Sn content layers, with therefore a lower density of misfit dislocations to deal with. In order to investigate the role of band gap directness on the lasing characteristics, once defects were removed, we explored Sn contents in the 7 \% to 10 \% range. We thus probed $\Delta E_{L-\Gamma}$ close to 0 up to 100 meV. We have obtained a drastic reduction of lasing thresholds in GeSn microdisk cavities, as compared to the literature, even for the lowest Sn content sample which had the smallest directness. A specific processing that suppressed interface defects in the active region of the cavity has most likely enabled such improvement. This highlights the key role of defects removal on lasing performances, a strategy which can be used, in the future, in structures with higher Sn contents and therefore higher band gap directness.

\section{Results and discussion}

\subsection{Fabrication}

 GeSn layers with various Sn contents were grown on Ge-VS on Si (001) substrates. Structural parameters such as residual strain and Sn content were quantified using X-Ray-Diffraction (XRD), Raman spectroscopy and Transmission Electron Microscopy (TEM) (see supporting information). Structural parameters are summarized in Table \ref{tbl:example}.

\begin{figure}[ht]
\centering
\includegraphics[width=\linewidth]{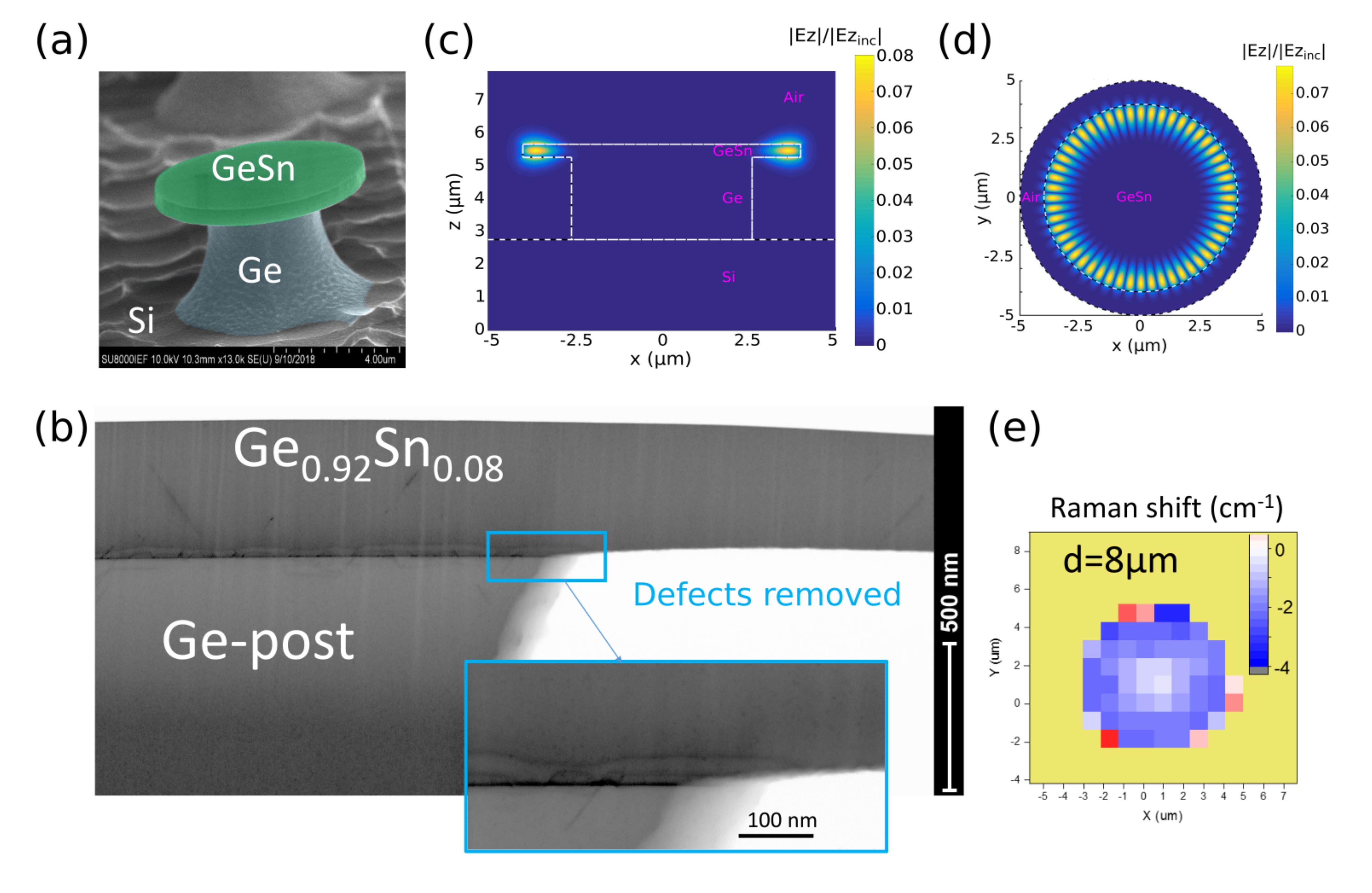}
\caption{ \textbf{a} 3D Scanning Electron Microscopy image of a GeSn microdisk \textbf{b} Bright Field (BF) TEM image of a GeSn microdisk. The lamella was milled by Focused Ion Beam through a microdisk with a \SI{10 }{\micro\meter} diameter \textbf{c} Electromagnetic field $|E_z|/|E_{z-inc}| $ of a TM whispering gallery mode (L=26, N=1) with an energy of 0.5352 eV in the Oxz plane. \textbf{d} Same as c) in the Oxy plane, the dotted circle shows the edges of the Ge pillar. \textbf{e} 2D map of Raman shift measured on a \SI{8 }{\micro\meter} diameter microdisk. The Raman position of TO vibration mode of the as-grown layer is taken as a reference.} 
\label{fig:Fab}
\end{figure}

\begin{table}
  \caption{Structural parameter of the GeSn layers after growth}
  \label{tbl:example}
  \begin{tabular}{lll}
    \hline
    \% [Sn] & Strain & Nominal thickness\\
    \hline
      7 & -0.3 \% & 500 nm \\
     8.1 & -0.37 \% & 500 nm\\
      10.5 & -0.53 \% & 500 nm\\
    \hline
  \end{tabular}
\end{table}

Microdisk cavities were etched into the as-grown layers using the standard processing tools described in the supporting information (SI), with diameters ranging from 4 to \SI{10 }{\micro\meter} (Fig. \ref{fig:Fab}-a). Here all microdisks were underetched by typically \SI{1.5 }{\micro\meter} from the edges, leaving a very narrow Ge pillar in the central part for disks with a diameter of \SI{4 }{\micro\meter}.\cite{herth_fast_2019} As compared to other reports,\cite{stange_optically_2016,reboud_optically_2017} we used an etching recipe with SF$_6$ instead of CF$_4$ gas. The latter yielded a very high selective etching of Ge over GeSn,\cite{gupta_highly_2013} and was thus selected as the best option to fabricate suspended GeSn microdisks in the literature.\cite{du_study_2019} However, such an etching chemistry does not enable to etch the bottom defective interface in GeSn as discussed in ref. \cite{stange_optically_2016}. On the other hand, the SF$_6$ gas is less selective and enables, as discussed in the following, to remove the defective interface from the active region of the GeSn microdisks. Figure \ref{fig:Fab}-b shows a bright field TEM image of a GeSn microdisk layer with a diameter of \SI{10 }{\micro\meter} (more details in the SI). One can see that the as-grown GeSn layer presents a high density of stacking faults and misfit dislocations near the GeSn/Ge interface. Such defects recombine in the first 60 nm by forming dislocation loops parallel to the interface. A lower density of stacking fault segments extend more in depth in the layer with an angle of \SI{54}{\degree} but without necessarily crossing the whole layer. The Ge etching recipe used to fabricate the pillar was not infinitely selective over GeSn. This led to a reduced thickness on its bottom side (see SI) as the top surface was protected by the photoresist that defines the pattern. More importantly, the partial etch of the bottom part of the GeSn film resulted in a total removal of the dense misfit dislocation network at the interface. Even in poorly etched areas, i.e. near the Ge pillar, (inset of Fig. \ref{fig:Fab}-b), the layer still had defects removed. Additionally, the stacking fault segments which extended more in depth disappeared from the suspended GeSn area. Both features suggest that additional curing might have occurred in GeSn (see SI). During the Ge etch, the initial GeSn/Ge interface becomes a free GeSn surface which can reconstruct. Curing and enhancement of defect mobility may have been activated by local heating, induced by the plasma etching chemistry. The fact that the GeSn is increasingly suspended in the air during the underetching of Ge minimizes thermal dissipation to the substrate. The current analysis was performed on the sample with an intermediate Sn content of 8 \%. Measurements of the GeSn layer thickness at the microdisk edges for Sn contents of 7 \% and 10.5 \% showed a systematic reduction of this defective region (see SI). One could thus assume that the same defect removal occurred over the whole Sn content range probed in the current work. We emphasize that defects were removed in the most relevant region of the cavity, given that laser emission comes from whispering gallery modes resonances (WGM) (Fig. \ref{fig:Fab}-c-d). The suppression of dislocations in the outer-parts of the GeSn microdisk should have a positive impact on carrier recombination dynamics, optical losses, and thus on material gain and threshold. 

A first analysis by Raman spectroscopy of microdisk strain showed that strain relaxation occurred in suspended parts of the layer. Figure \ref{fig:Fab}-e is a 2D-surface map of Raman shift measured using the Raman spectral position of the as-grown layer as a reference. The central area of the microdisk above the Ge pillar has quasi-unrelaxed strain, here for a microdisk with a \SI{8 }{\micro\meter} diameter. Strain relaxation, characterized by a Raman red-shift, occurs in the suspended part of the layer, which has an optimized overlap with WGMs. The relaxation of compressive strain enables to increase the band gap directness parameter and promote optical gain\cite{stange_optically_2016,gupta_achieving_2013} in this defect-free region. Further Raman analysis was performed on the microdisks to assess the strain relaxation in small microdisk diameters of around \SI{4 }{ \micro\meter}, as studied in photoluminescence (PL). A homogeneous strain distribution along the diameter was obtained in those cases (see SI).

\subsection{Optical analysis}

\begin{figure}[H]
\centering
\includegraphics[width=\linewidth]{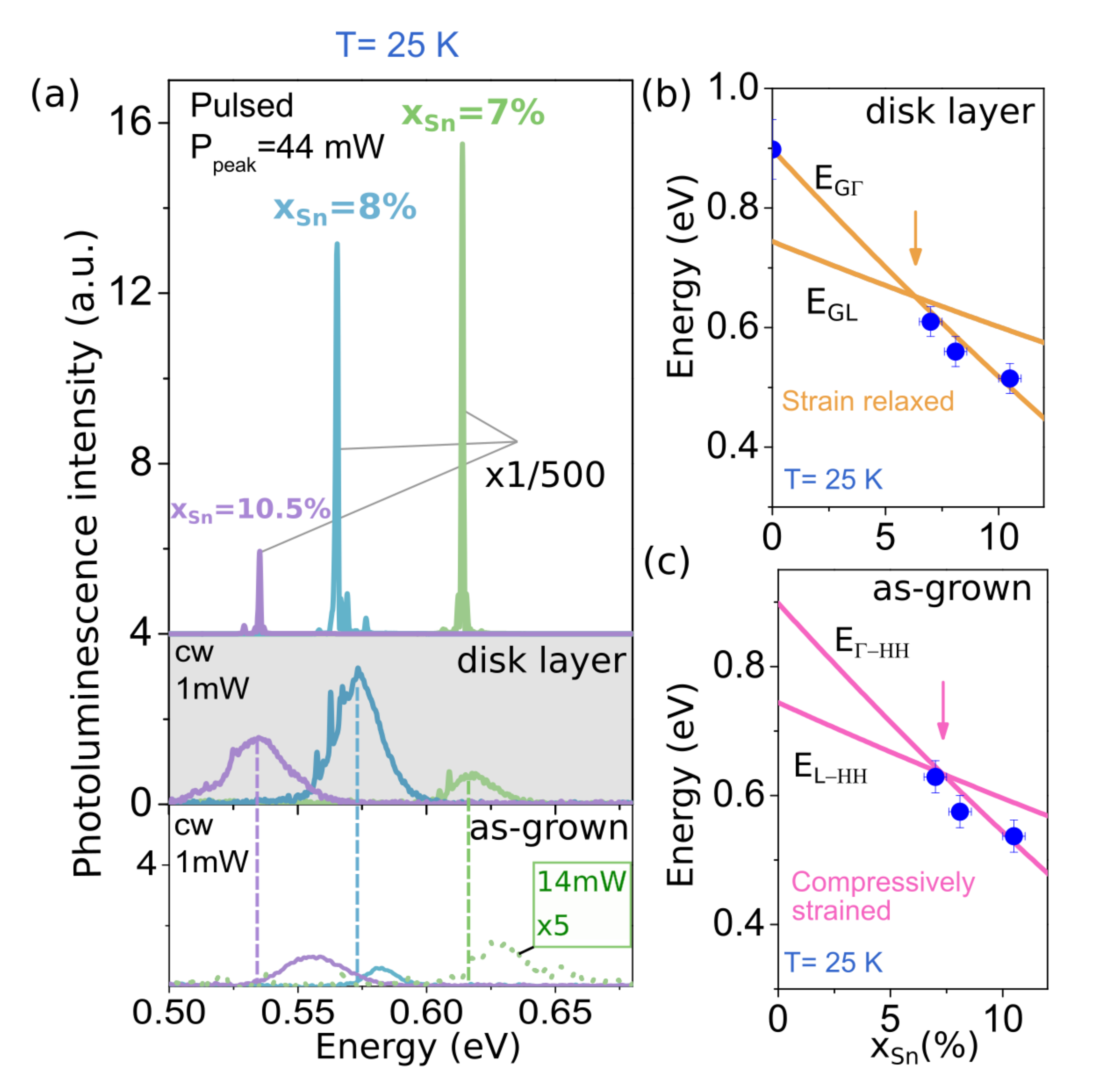}
\caption{ \textbf{a}-bottom, PL spectra from as-grown GeSn layers under \SI{1 }{\milli\watt} cw pump power (\SI{0.9 }{\kilo\watt\per\square\centi\meter} power density), except for the $x_{Sn}=7 \%$ layer pumped with a \SI{14 }{\milli\watt} pump power (\SI{12 }{\kilo\watt\per\square\centi\meter} power density) and a signal multiplied by a factor of 5. Middle, PL spectra from layers patterned into \SI{4 }{\micro\meter} diameter microdisks, under cw excitation of \SI{1 }{\milli\watt}. Top, same as middle but excitation is turned into a pulsed mode with a \SI{44 }{\milli\watt} peak power, (\SI{38 }{\kilo\watt\per\square\centi\meter} power density). Spectra has been offset and divided by a factor of 500. \textbf{b} Experimental values of fundamental band gap obtained from PL spectra under cw pumping of microdisks. Orange lines correspond to modeled direct and indirect band gap energies $E_{G\Gamma}$ and $E_{GL}$ of relaxed GeSn layers with bowing parameters $b_{L}=0.89$ eV and $b_{\Gamma}=2.77$ eV. The arrow shows the indirect-direct bandgap crossover. \textbf{c} same as (b) but for as-grown layers assuming 30 \% of residual compressive strain due to the lattice parameter mismatch (between GeSn and Ge).} 
\label{fig:Fig1}
\end{figure}

The PL analysis of GeSn layers with various Sn contents was performed at 25 K under \SI{1550}{\nano\meter} wavelength optical pumping. Spectra are shown in Figure \ref{fig:Fig1}-a as follows : bottom: PL spectra under continuous wave (cw) pumping on as-grown layers, middle: after patterning of the layers into small \SI{4 }{\micro\meter} diameter microdisks under cw excitation, and top: under 3.5 ns duration pulsed excitation with a 25 MHz repetition rate. 
As discussed below, the combination of 7 \% of Sn with 0.3 \% of compressive strain should result in an indirect band gap alignment of the band structure. As shown in Figure \ref{fig:Fig1}-a bottom, the layer with 7 \% Sn exhibits a much lower PL signal as compared to the other samples with higher Sn contents. The signal was indeed too low to be detected under \SI{1 }{\milli\watt} excitation power. We thus had to increase the excitation power up to \SI{14 }{\milli\watt} and multiply the obtained signal intensity by a factor of 5 to compare its amplitude with the signals for $x_{Sn}=8 \%$ and 10.5 \% under \SI{1}{\milli\watt} pump power excitation. 

After patterning the layers into microdisks, we observed clear red-shifts in PL spectra due to the full relaxation of the residual compressive stress by underetching. Note that the spectra from \SI{4 }{\micro\meter} diameter microdisks are shown in Figure \ref{fig:Fig1}-a using the same scale as for PL spectra on as-grown layers to facilitate comparison.

The band gaps were extracted from the PL spectra under cw excitation. They are plotted in Figures \ref{fig:Fig1}-b and -c for respectively the microdisks and as-grown layers. In the same figure, we have plotted the theoretical band gap of GeSn alloys assuming 30 \% (see SI) of residual compressive strain for the as-grown layers (Figure \ref{fig:Fig1}-c). Meanwhile, the strain is assumed to be completely relaxed in microdisks with very narrow Ge pedestals (Figure \ref{fig:Fig1}-b) as evidenced by Raman spectroscopy (see the SI). The experimental energy dependencies on Sn content are properly fitted with an empirical quadratic law $E_{\Gamma,L}(x_{Sn})=E_{\Gamma,L}^{Ge}\times({1-x_{Sn}})+ E_{\Gamma,L}^{Sn}\times x_{Sn}-b_{\Gamma,L}(1-x_{Sn})x_{Sn}$. The following valley energies were fixed: $E_{\Gamma}^{Sn}=-0.408$ \SI{}{\electronvolt} and $E_{L}^{Sn}=0.12$ \SI{}{\electronvolt}, $E_{\Gamma}^{Ge}=0.898$ \SI{}{\electronvolt}, $E_{L}^{Ge}=0.744$ \SI{}{\electronvolt} while $b_{L}=0.89$ \SI{}{\electronvolt} according to ref.\cite{Rainko2018a}. As we have achieved lasing for microdisks with Sn contents above 7 \%, we therefore assume that the microdisk emission stems from direct transitions. The experimental band gap dependence on Sn content shown in Figure \ref{fig:Fig1}-b was thus fitted with $E_{\Gamma}(x_{Sn})$ by adjusting the bowing parameter and adopting $b_{\Gamma}=2.77$ \SI{}{\electronvolt}, a value close to the $2.46$ \SI{}{\electronvolt} energy reported in ref.\cite{bertrand_experimental_2019}. In previous reports, for instance in ref.\cite{Rainko2018a,rainko_investigation_2018}, where $b_{\Gamma}=2.24$ \SI{}{\electronvolt} and $b_{L}=0.89$ \SI{}{\electronvolt} were used, the indirect-direct band gap crossover was expected to occur at around 8 \% of Sn, as also reported in ref. \cite{wirths_lasing_2015,BowingMenendez,GeSnKouvekatisa,grzybowski_next_2012,menendez_materials_2019}. Here, it occurred at 6.4 \%, in good agreement with the modeled one for relaxed GeSn alloys.\cite{dutt_theoretical_2013} In ref.\cite{menendez_materials_2019,BowingMenendez}, a composition-dependent law was proposed: $b_{\Gamma}=2.66-0.54x_{Sn}$, e.g. around \SI{2.61 }{\electronvolt} for the 7-10\% Sn content range considered here and in satisfying agreement with the \SI{2.77 }{\electronvolt} value that we found, even if it remains lower. In ref.\cite{menendez_materials_2019,BowingMenendez}, the crossover is however found also for an higher Sn content, e.g. 8\%, than in this work. The differences in crossover values may stem from the presence of residual compressive strain in the layers used to determine it and from a higher value for $b_{L}$, in the range of \SI{1 }{\electronvolt} in ref.\cite{menendez_materials_2019,BowingMenendez}. Here the crossover value comes from fits of data on suspended, strain-free layers. The value of 0.89 eV for $b_{L}$ is consistent with our experimental results, i.e. that the relaxed layer with 7\% Sn content has a direct band gap at 25 K and an indirect band gap when compressively strained by 0.3\%. 

The latter bowing parameters were subsequently used to calculate the band gap of as-grown layers (see SI), taking into account the impact of residual compressive strain on the band structure (30 \% of the residual strain due to lattice parameter mismatch for all Sn contents). Since the compressive strain lifts the degeneracy of the valence band with the heavy-hole (HH) band being the highest in energy, we calculate the direct $E_{\Gamma-HH}$ and indirect $E_{L-HH}$ gap energies and compare them with data from as-grown layers (Figure \ref{fig:Fig1}-c). Figure \ref{fig:Fig1}-c indicates that the combination of 7 \% of Sn with 0.3 \% of compressive strain should result in an indirect band gap alignment.

We have observed, for all Sn contents, an enhancement of the PL emission for patterned, relaxed, layers as compared to still compressively-strained as-grown ones as seen in Figure \ref{fig:Fig1}-a middle. The enhancement depends on the Sn content (enhancement factors of 1.5 and 5 for 10.5 \% and 8 \% of Sn to be compared with two orders of magnitude for 7 \% of Sn). Several factors can explain the microdisk PL amplitude change as compared to as-grown layers : i) obviously, improved radiative efficiency because of defect removal from the bottom of GeSn layers ii) the change of radiation pattern of the overall emission in patterned layers, and iii) the increase of the band structure directness, favoring an electron population in the $\Gamma$ valley after optical excitation. Carriers in the $\Gamma$ valley have a higher radiative recombination rate at zone center, i.e. a higher PL efficiency, while carriers in the \textit{L} valley do not contribute significantly to the PL signal. 

For the GeSn layer with $x_{Sn}=7$ \%, the PL enhancement is very strong since the band gap changes from indirect to direct. We estimate $\Delta E_{L-\Gamma}$ to be \SI{-8 } {\milli\electronvolt} in the as-grown layer, i.e. the quasi-totality of electrons are in the L-valley at low temperature and thus the PL signal stems from weak indirect transitions. Meanwhile, $\Delta E_{L-\Gamma}=16$\SI{} {\milli\electronvolt} in the strain-free microdisk layer, resulting in a strong $N_{\Gamma}$ population enhancement and the contribution of direct transitions to the PL signal. A clear signature of direct band alignment in the strain-free sample with $x_{Sn}=7$ \% is the observation of lasing when the microdisk is optically pumped at higher powers using pulsed excitation (Figure \ref{fig:Fig1}-a - see discussion in the next paragraph).

For the $x_{Sn}=8$ \% sample, the $\Delta E_{L-\Gamma}$ splitting energy increases from \SI{11}{\milli\electronvolt} up to \SI{39}{\milli\electronvolt} resulting as well in a significant increase of the $N_{\Gamma}$ population. The change of directness for this sample also explains the higher PL amplitude.

For the $x_{Sn}=10.5 \%$ sample, the directness is expected to change from \SI{51 }{\milli\electronvolt} in the as-grown layer with residual compressive strain to \SI{94 } {\milli\electronvolt} in the strain-relaxed microdisk. The low temperature electron population $N_{\Gamma}$ in the $\Gamma$ valley is expected to be equivalent in both cases. The directness increase is thus not the parameter explaining the PL signal enhancement at low temperature. It is rather, in this case, the scattering of emitted light at the microdisk edges and the interface defect removal that predominate. Note that electromagnetic simulation (with Fourier-Bessel modal method)\cite{bonod_differential_2005} show a lower light absorption in the GeSn microdisks, by a factor of 0.8, as compared to as-grown layers during the optical pumping. While the microresonator absorbs light more efficiently, the smaller illuminated volume of the microdisk (with \SI{4 }{ \micro\meter} diameter only, while the excitation spot size was \SI{12 }{\micro\meter}) limits the radiative recombination rate.

\subsection{Low temperature lasing}

\begin{figure}[H]
\centering
\includegraphics[width=\linewidth]{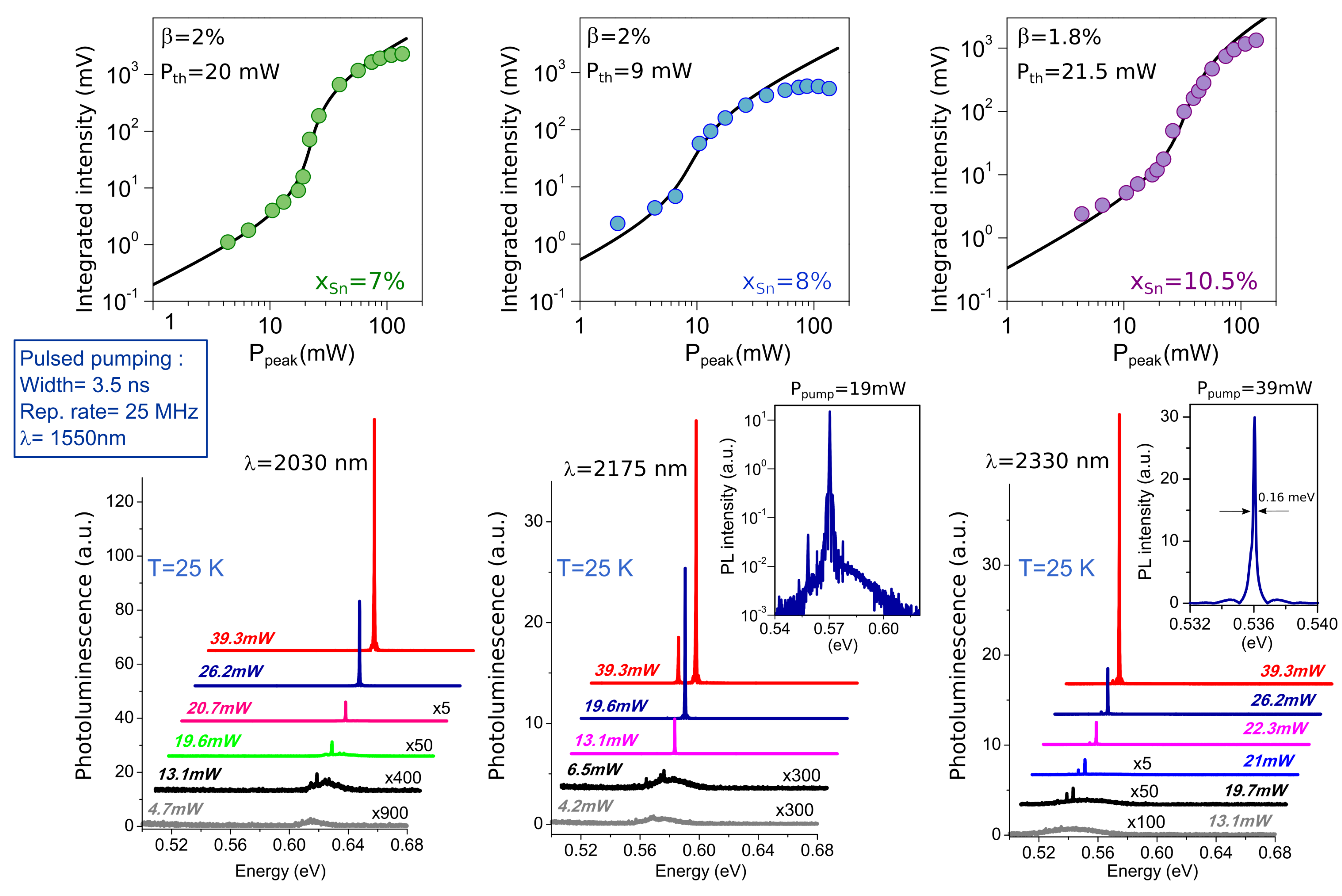}
\caption{(Bottom) Pulsed peak power dependence of emission spectra for \SI{4 }{\micro\meter} diameter microdisks with Sn contents between 7 \% to 10.5 \% with corresponding L-L curves at \SI{25 }{\kelvin} (Top). The continuous curves in the figures on top are calculated curves using the laser rate equations with the indicated thresholds and beta factors. The insets show typical spectra above lasing threshold.} 
\label{fig:LasePulse}
\end{figure}

Low temperature, at 25 K, PL measurements were performed, for each Sn content, on \SI{4 }{\micro\meter} diameter microdisks. The power dependence and the measured PL spectra as well as the integrated intensity (L-L) curves are shown in Figure \ref{fig:LasePulse}. These measurements were performed in a pulsed excitation regime with a \SI{1.55 }{\micro\meter} wavelength laser, with \SI{3.5 }{\nano\second} long pulses and a repetition rate of 25 MHz. The laser beam was focused into a \SI{12 }{\micro\meter} diameter spot on the sample surface. Under these conditions, the pump power density can be obtained by multiplying the incident power on the sample surface by a \SI{0.88d6 }{\per\square\centi\meter} factor. For all samples, we reached a laser emission regime, characterized by an abrupt transition from broad and weak to intense and narrow emission lines at a well-defined pump power threshold. Above threshold, as typically shown in the inset of Figure \ref{fig:LasePulse} for the 8 \% sample, a single lasing mode dominates the spontaneous emission background by typically 3 orders of magnitude. Furthermore, the linewidth of the laser mode is typically in the \SI{100 }{\micro\electronvolt} range against meV in previously reported GeSn lasers. We emphasize that this small linewidth can be observed here thanks to the experimental conditions and the low lasing threshold. Indeed, high photon flux yielded by high duty cycle excitation enables high-resolution analysis down to \SI{60 } {\micro\electronvolt}. In previous studies, the very high lasing thresholds required the use of low optical pumping duty cycles in the 10$^{-5}$ to 10$^{-2}$ range, to reach the requested very high pump power densities while avoiding sample heating and damaging. In the present work, the lower lasing thresholds enabled us to use only 10$^{-1}$ of duty cycle and thus a higher photon flux without any microdisk damages. 

The L-L curves for all microdisks show an S-shape and can be fitted using the laser rate equations given in ref.\cite{rosencher_vinter_2002}. We extracted from the L-L curves the thresholds and the spontaneous emission beta factors, as given in the inset. The beta factors were obtained by integrating the whole emission spectrum to enable comparison with previous reports on GeSn microdisk lasers.\cite{stange_optically_2016,reboud_optically_2017, Thai:18}. This strategy differs from the standard approach where only the power in the lasing mode is considered to plot the L-L curves. Beta factors, around 2 \%, were uncorrelated with the Sn content. Such a value is much higher than values usually measured for microdisk cavities in the literature\cite{selles_deep-uv_2016}. We emphasize that when the emission is collected from the top surface, as done here, one has a strong contribution from spontaneous emission over the lasing whispering gallery mode, and the plotted L-L curve is less representative of the laser dynamics \cite{mohideen_gaas/algaas_1994}. Only a small fraction, of typically a few \% of the WGMs which radiate preferentially along the disk plane, is collected by the objective. Moreover the spontaneous emission, which is preferentially collected from the top surface since it radiates vertically, has a much broader spectral distribution than the WGM lasing mode.

The thresholds were determined as \SI{17.6 }{\kilo\watt\per\square\centi\meter}, \SI{8 }{\kilo\watt\per\square\centi\meter} and \SI{19 }{\kilo\watt\per\square\centi\meter} for microdisks with 7 \%, 8 \% and 10.5 \% of Sn. Those thresholds are in the \SI{10 }{\kilo\watt\per\square\centi\meter} range, to be compared with previous thresholds in higher Sn content lasers, typically a few hundreds of \SI{}{\kilo\watt\per\square\centi\meter}. More specifically, the lasing threshold for the GeSn 8 \% is much lower than those in previous reports,\cite{stange_optically_2016,Laser_GeSn_Arkansas} i.e. \SI{130 }{\kilo\watt\per\square\centi\meter} for a microdisk with an equivalent Sn content. These values are also lower than those obtained in a specifically-designed multi-quantum well structure (\SI{40 }{\kilo\watt\per\square\centi\meter}) that should in principle have a reduced threshold, through the quantization of electronic state energies, and separation of optically active media away from the defective regions \cite{StangeMQWlaser}. 

\begin{figure}[h]
\centering
\includegraphics[width=\linewidth]{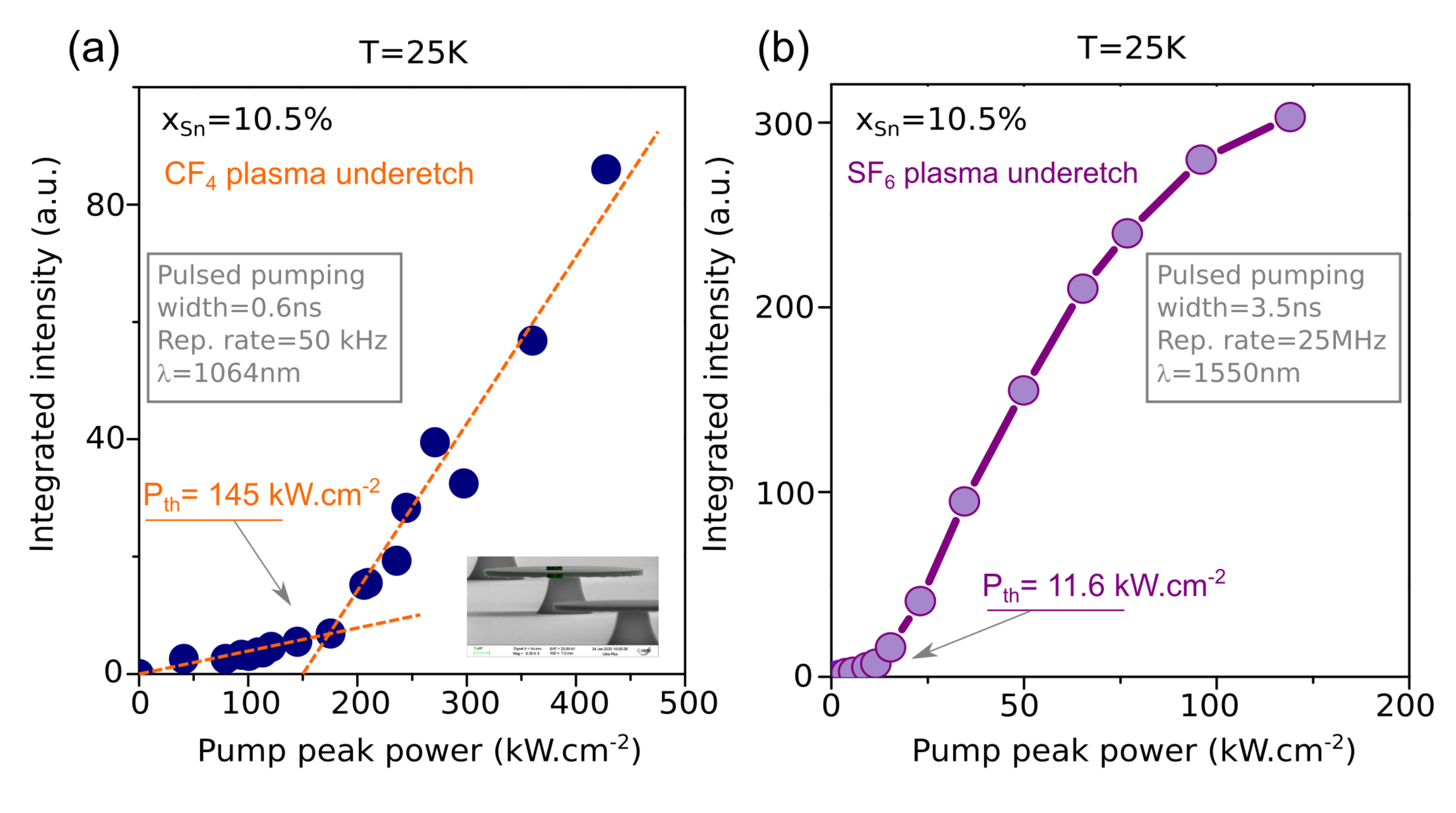}
\caption{ \textbf{a} L-L curve obtained at 25 K from a \SI{10 }{\micro\meter} diameter microdisk with 10.5 \% of Sn underetched with a CF$_4$ plasma. The dashed lines are drawn to guide the eyes. The inset shows a SEM image of such microdisks \textbf{b} L-L curve obtained at 25 K from a \SI{8 }{\micro\meter} diameter microdisk fabricated from the same epitaxial layer with 10.5 \% of Sn but underetched with a SF$_6$ plasma.} 
\label{fig:Fig3bisavecCF4}
\end{figure}

The use of SF$_6$ gas for the microdisk underetching is a key asset to reach such low thresholds. Figure \ref{fig:Fig3bisavecCF4} shows the lasing characteristics of a microdisk fabricated with the same sample with 10.5 \% of Sn but using CF$_4$ as underetching gas. The threshold is around \SI{145 }{\kilo\watt\per\square\centi\meter}, i.e. in the same range as mentioned above from literature data. Such high excitation range required to change the optical pumping scheme to avoid sample heating and damaging. Shorter pulse duration (\SI{0.6 }{\nano\second}) and lower repetition rate (\SI{50 }{\kilo\hertz}) were used as indicated on figure \ref{fig:Fig3bisavecCF4}-a. As a comparison, we show the lasing characteristics (more data in the SI) of a microdisk of an equivalent size underetched by the SF$_6$ plasma. In this case the lasing threshold is around \SI{11.6 }{\kilo\watt\per\square\centi\meter}. This indicates the prominent role played by microdisk processing on the laser performances, beyond the arguments on Sn content and band structure directness. Starting from the same active region, a reduction by more than one order of magnitude of the threshold can be obtained with an appropriate processing. The removal of the dense array of defects from the active region is obviously one key feature.

\subsection{Temperature dependence}

\begin{figure}[H]
\centering
\includegraphics[width=\linewidth]{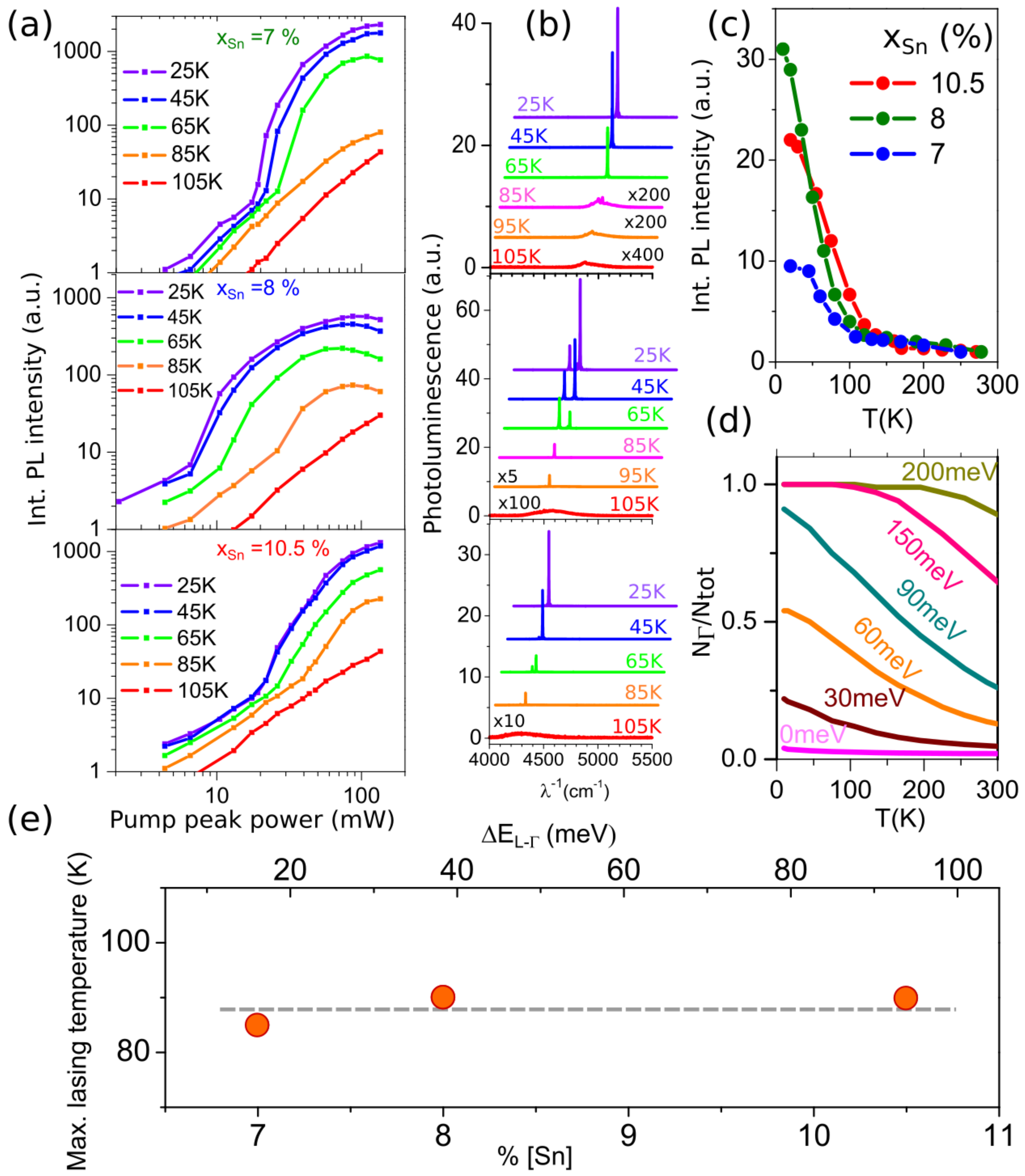}
\caption{ \textbf{a} L-L curves at temperatures from \SI{25}{\kelvin} to \SI{105}{\kelvin}, for GeSn microdisks with a \SI{4 }{\micro\meter} diameter and Sn contents x$_{Sn}$ of 7 \%, 8.1 \% and 10.5 \%. No lasing is observed at temperatures above \SI{100}{\kelvin}. \textbf{b} Emission spectra above lasing threshold for temperatures varying from \SI{25}{\kelvin} to \SI{105}{\kelvin}.\textbf{c} Integrated PL signal from microdisks with different Sn contents as a function of temperature measured under cw excitation below lasing threshold.\textbf{d} Calculated electron population ratio in the $\Gamma$ valley as a function of temperature for different conduction band barrier energies $\Delta E_{L-\Gamma}=E_L-E_{\Gamma}$.\textbf{e} Maximal lasing temperature as a function of investigated \% of Sn and corresponding directness parameter.} 
\label{fig:Fig3}
\end{figure}

We have studied the lasing temperature dependence for different Sn content microdisks to understand the influence of band gap directness on maximum lasing temperature. The L-L curves measured for various Sn content microdisks and different temperatures are plotted in Figure \ref{fig:Fig3}-a. The emission spectra obtained with pump powers above thresholds are shown in Figure \ref{fig:Fig3}-b. Additional measurements with 5 K temperature steps (presented in the SI) show that the sample with 7 \% of Sn can even support lasing at temperatures up to 80-85 K, like samples with Sn contents of 8 \% and 10.5 \% which support lasing up to 85-95 K. The general trend in the literature is that higher Sn contents result in higher maximum lasing temperature. The main reason is a reduction of the carrier scattering rate from to $\Gamma$ to L valleys due to an increase of the energy barrier $\Delta E_{L-\Gamma}$, enabling an efficient injection of electrons in the $\Gamma$ valley. 

 In our experiment, we observe very similar maximum lasing temperatures for microdisks, with a maximum temperature increase of only 10-15 K for Sn contents between 7 \% and 10.5 \%, while $\Delta E_{L-\Gamma}=E_L-E_{\Gamma}$ is expected to vary significantly, from 16 meV for 7 \% of Sn up to 94 meV for 10.5 \% of Sn (Figure \ref{fig:Fig3}-e). Since the quenching of lasing with increasing temperature can be related to the quenching of population in the $\Gamma$ valley, we have studied the microdisk PL below threshold under cw pumping as a function of temperature. The PL signal is sharply quenched at temperature above 100 K in a similar way for the various Sn content microdisks (Figure \ref{fig:Fig3}-c). Experimentally, the quenching of PL signal is therefore independent of $\Delta E_{L-\Gamma}$. To better quantify the thermal activation of $\Gamma$-L scattering, we have calculated the equilibrium electron population ratio $N_{\Gamma}\over{(N_{\Gamma}+N_L)}$ as a function of temperature for several values of the splitting energy between both bands $\Delta E_{L-\Gamma}$. Figure \ref{fig:Fig3}-d shows the predicted electron distribution rate in the $\Gamma$ valley as a function of temperature for different $\Delta E_{L-\Gamma}$ from 0 up to 200 \SI{}{\milli\electronvolt} and fixed total carrier density $(N_{\Gamma}+N_L)$ of $10^{18}$ \SI{}{\per\cubic\centi\meter}. The $\Gamma$-valley population varies slowly with temperature and does not show a quenching as sharp as the one observed in PL signal. It is thus reasonable to suppose that other parameters influence the maximum lasing temperature beside band structure parameters. 
 
 Most probably, the abrupt quenching of lasing and PL for all Sn contents investigated here can be associated with the activation of non-radiative processes as the temperature increases. Above 100 K, non-radiative processes dominate the radiative process and prevent lasing. A systematic study has recently evidenced the presence of vacancy defects in as-grown layers with Sn contents from 6 \% to 13 \%, corresponding to the range probed here.\cite{AssaliVacancy} These defects might be due to the low temperature required to grow metastable GeSn layers above the Sn solubility limit. Vacancies result in a p-type doping of the layers, and thus in optical losses when activated by a temperature increase. Additionally, p-type defects are usually known to shorten coherence lifetimes due to valley scattering. It increases the homogeneous broadening of oscillator strength and the optical gain is therefore weakened.\cite{ghrib_all-around_2015}

\subsection{Discussion}
Microdisks with different diameters and pillar diameters were investigated as well (See SI). For microdisks with an undercut of \SI{1.5 }{\micro\meter}, we have obtained reproducible lasing threshold values for diameters of 4-\SI{5 }{\micro\meter} (See SI), \SI{17.6 }{\kilo\watt\per\square\centi\meter} for $x_{Sn}=7 \%$, \SI{8 }{\kilo\watt\per\square\centi\meter} for $x_{Sn}=8 \%$ and \SI{8.9 }{\kilo\watt\per\square\centi\meter} for $x_{Sn}=10.5 \%$. The thresholds were reduced when the Sn content increased from $x_{Sn}=7 \%$ to $x_{Sn}=8 \%$ and remained quasi-unchanged for higher Sn contents.

\begin{figure}[H]
\centering
\includegraphics[width=\linewidth]{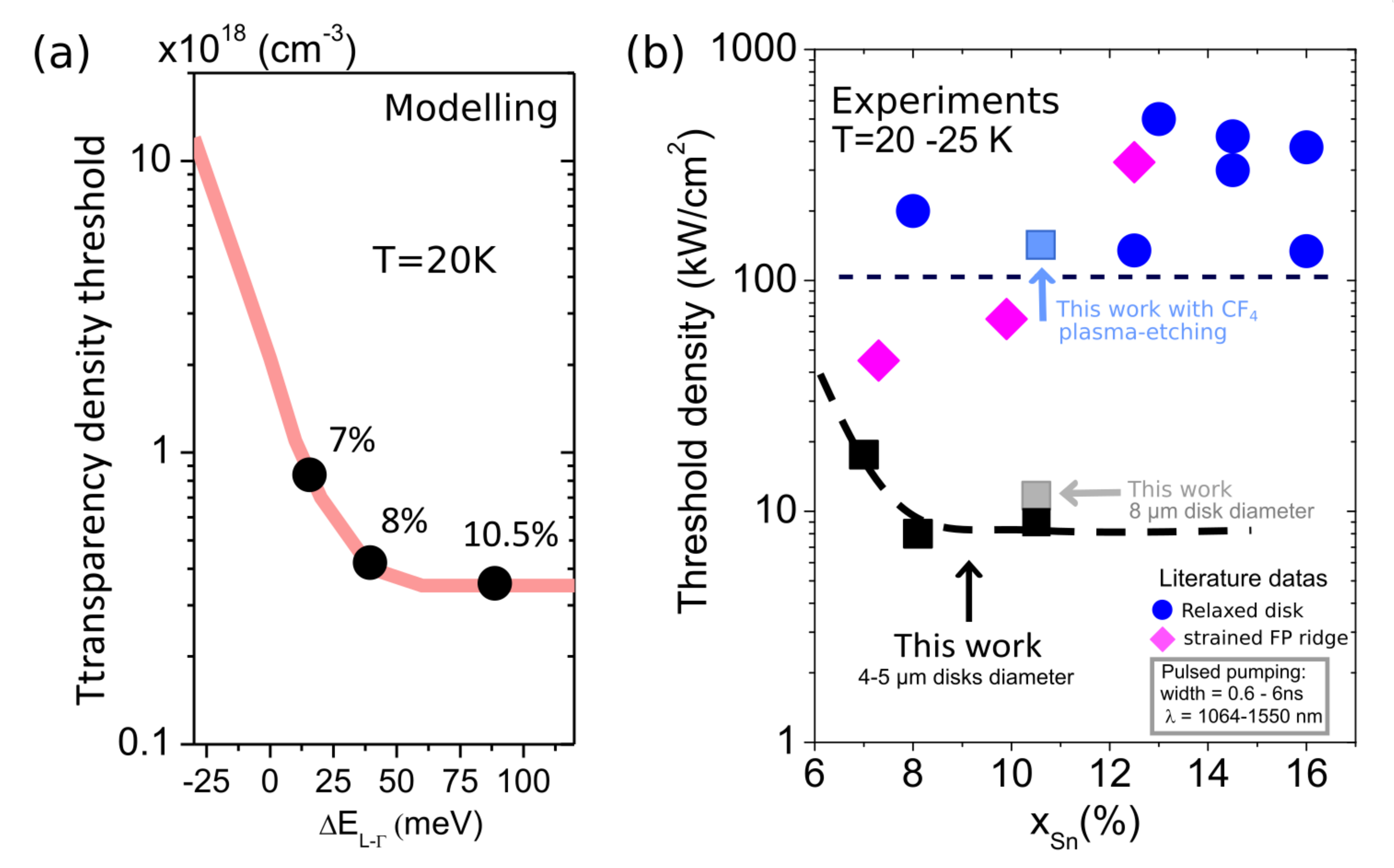}
\caption{ \textbf{a} Carrier density threshold to reach transparency as a function of the directness parameter $\Delta E_{L-\Gamma}$ at 20 K in strain-free material. The black circles show the corresponding relaxed GeSn microdisks with the Sn contents probed in this work \textbf{b} Review of low temperature lasing thresholds in GeSn devices as a function of the Sn content in the active media (FP stands for Fabry-Perot). We show only threshold below 25 K. Relaxed GeSn disk cavities are shown as blue circles,\cite{reboud_optically_2017,stange_optically_2016,Thai:18} while results obtained with compressive GeSn ridge waveguides appear as pink diamonds \cite{LaserArk180K,Laser_GeSn_Arkansas,wirths_lasing_2015}, including complex multilayers. More data and details on experimental conditions used for previous works can be found in the comparison table reported in ref.\cite{du_study_2019}. Black, grey and light blue squares are current values. The dashed line is drawn to show the trend.} 
\label{fig:Fig6}
\end{figure}

The decrease of the lasing threshold when increasing the Sn content is similar to the one modeled for the carrier density threshold to obtain transparency as a function of $\Delta E_{L-\Gamma}$, as shown in Figure \ref{fig:Fig6}-a. 

The method used for the calculation is detailed in ref. \cite{ghrib_all-around_2015,boucaud_recent_2013,elbaz_ultra-low-threshold_2020}. On this figure we can see that, at low temperature, the carrier density threshold to reach transparency decreases very rapidly at the indirect-direct crossover and stabilizes typically for $\Delta E_{L-\Gamma}$ above \SI{30}{\milli\electronvolt}, i.e. for $x_{Sn}$ above 8 \%. Here, the modeling predicts a threshold reduction by typically a factor 2 from $x_{Sn}=7 \%$ to $x_{Sn}=8 \%$, which can instructively be compared with the threshold reduction observed experimentally for these layers, from \SI{17.6 }{\kilo\watt\per\square\centi\meter} for $x_{Sn}=7 \%$ down to \SI{8 }{\kilo\watt\per\square\centi\meter} for $x_{Sn}=8 \%$. The present study shows that lasing thresholds for Sn contents higher than 8 \% should stabilize slightly below \SI{10 }{\kilo\watt\per\square\centi\meter}, while previous reports, even with different cavity designs, showed a very wide range of thresholds.

We note that the measured thresholds can depend on many factors, like the microdisk thickness, the pillar sizes, the laser cavity geometries as well as the excitation conditions (pump wavelength, duty cycle, and repetition rate). The data shown on Figure \ref{fig:Fig6}-b use the comparison data reported in ref.\cite{du_study_2019} with a very wide range of pumping conditions and cavity designs, which are recalled on figure \ref{fig:Fig6}-b were considered. In this work, the thicknesses are very similar to those reported in ref.\cite{du_study_2019}. The optical pumping conditions are also similar, only the duty cycle is much higher here. Note that it is not favorable, since as discussed in the SI, a higher duty cycle induces heating of the gain media as compared the low duty cycle used in previous reports. All reported thresholds are given in terms of peak power densities which govern the injected carrier densities, independently of the duty cycle.
Our measurements suggest that there is no clear threshold dependence on Sn contents above 8 \%, i.e. on band structure directness. We suggest that the limiting factors previously reported were not band structure issues,\cite{Laser_GeSn_Arkansas, LaserArk180K, wirths_lasing_2015, Dou2018a, reboud_optically_2017, Thai:18,stange_optically_2016} but material and processing issues, that have been partly solved here thanks to the removal of the defective GeSn/Ge interface. We recall that the lasing threshold depends on the carrier density needed to reach transparency, i.e. directly dependent on the density of states, the cavity characteristics implying the dynamic gain and quality factor, and the radiative and non-radiative lifetimes. From our modeling of carrier density thresholds and measured lasing power density thresholds, one can extract an effective carrier lifetime following the generation-recombination balance law $N=I\tau/(h\nu d)$, where $I$ is the absorbed power density, $\tau$ the recombination lifetime, $h\nu$ the energy of absorbed photon. d is the layer thickness where the generated carrier density N is distributed. The incident power is multiplied by 0.65 to account for the microdisk surface reflectivity at 1550 nm wavelength. A pumping power of typically \SI{10 }{\kilo\watt\per\square\centi\meter} is found necessary to reach lasing. One can assume that lasing in the microdisks cavities can be reached for carrier densities of roughly 4 times the transparency thresholds. \cite{elbaz_ultra-low-threshold_2020} This corresponds to an injected carrier density of typically \SI{1.6d18}{\per\cubic\centi\meter} given that, as obtained from Fig. \ref{fig:Fig6}-a, the transparency threshold is obtained for carrier densities of typically \SI{4d17}{\per\cubic\centi\meter}. The extracted value for $\tau$ is then around 1.3 ns. This value is found in good agreement with the one obtained in a previous report \cite{elbaz_ultra-low-threshold_2020}, where the microdisk layers were made free from the interface defects, after they were transferred on a host Si-substrate. We postulate that the significant reduction in threshold observed in this work is associated with an increased carrier lifetime, as a consequence of the interface defects removal.
In ref. \cite{elbaz_ultra-low-threshold_2020}, it was shown that a bonding procedure can lead to a GeSnOI stack. In this approach, one can subsequently remove the defects entirely close to the surface, over the whole wafer area, by simply etching it and obtain layers free from the interface defects like in the present work. In this configuration, one can design various defect-free laser cavities, like Fabry Perot ridge, since there is no need to make additional underetching to remove the defects.

We have reached a maximum lasing temperature of around 95 K with Sn contents lower than or equal to 10.5 \%. Higher maximum lasing temperatures were reached with higher Sn content layers and therefore higher band gap directness.\cite{reboud_optically_2017,LaserArk180K} We however found that the maximum lasing temperature depended very weakly on the Sn content, at least in the range probed here, and thus on band gap directness.

We emphasize that an increased directness can also be obtained by applying tensile strain to the alloys instead of increasing their Sn content. These past years, several methods were indeed developed with the aim to change pure Ge into a direct band gap semiconductor and reach lasing with it. \cite{elbaz_germanium_2018,armand_pilon_lasing_2019,bao_low-threshold_2017} For Ge, 1.7 \% of biaxial tensile strain,\cite{el_kurdi_direct_2016,virgilio-rad-2013} and 4.9 \% uniaxial tensile strain were achieved. These methods could be used on GeSn alloys\cite{GeSnBridge} in order to increase their directness, notably for the reduced Sn contents probed here. Tensile strain even presents the advantage, over the increase of Sn content, of lifting the valence band degeneracy and thus reducing the density of states yielding lower gain threshold than in relaxed GeSn.\cite{el_kurdi_band_2010,Rainko2018a,elbaz_ultra-low-threshold_2020}

\section{Conclusion}
Low Sn content GeSn layers were used, despite their small band gap directness, to fabricate suspended microdisks exhibiting significantly reduced lasing thresholds as compared to previous reports in the literature. We used a specific processing step that led to the removal of the array of misfit dislocations at the GeSn/Ge interface in the active suspended area of the microdisk cavities. This was most likely the reason why we had a lasing threshold reduction by one order of magnitude as compared to the literature. We were able to confirm the influence of the band gap directness on lasing thresholds. Higher thresholds were obtained in microdisks with 7 \% of Sn, i.e. a lower band gap directness of only few \SI{}{\milli\electronvolt} as compared to samples with Sn contents of 8 \% and higher. We also found that in the explored range of 7 \%-10.5 \% range, the maximum lasing temperature depended only weakly on Sn content. Our results indicate that, beyond band gap directness, getting rid of non-radiative recombinations, including point defects, is mandatory to reach high temperature lasing.

\begin{acknowledgement}
E. Sakat and M. El Kurdi thank J.-P. Hugonin for fruitful discussions on electromagnetic simulations. This work was supported by the French RENATECH network, the French National Research Agency (Agence Nationale de la Recherche, ANR) through funding of the ELEGANTE project (ANR-17-CE24-0015). A. Elbaz was supported by ANRT through a CIFRE grant. Antonino Foti acknowledges funding within the ANR-16-CE09-0029-03 TIPTOP project.

\end{acknowledgement}

\begin{suppinfo}
Supplementary information file is available free of charge at http://pubs.acs.org. List of content : Material growth and characterization, Fabrication, Additional data with different microdisks sizes

\end{suppinfo}

%\bibliography{achemso-demo}

\providecommand{\latin}[1]{#1}
\makeatletter
\providecommand{\doi}
  {\begingroup\let\do\@makeother\dospecials
  \catcode`\{=1 \catcode`\}=2 \doi@aux}
\providecommand{\doi@aux}[1]{\endgroup\texttt{#1}}
\makeatother
\providecommand*\mcitethebibliography{\thebibliography}
\csname @ifundefined\endcsname{endmcitethebibliography}
  {\let\endmcitethebibliography\endthebibliography}{}

\end{document}